\documentstyle[12pt]{article}
\advance\textheight by \topskip
\begin{document}
\thispagestyle{empty}
\begin{flushright}
SU--ITP--94--32\\
IEM--FT--92/94\\
gr-qc/9409026\\
September 13, 1994
\end{flushright}
\vskip 2cm
\begin{center}
{\Large\bf Quantum fluctuations of Planck mass as\\
mutation mechanism in a theory of\\
evolution of the universe}\footnote{
Talk presented at the parallel session on Quantum Cosmology
of the Marcel Grossmann Meeting on Gravity, July 24-30, 1994.}\\
\vskip 1.5cm
{\bf Juan Garc\'{\i}a--Bellido}\footnote{
E-mail: bellido@slacvm.slac.stanford.edu}
\vskip 0.05cm
Department of Physics, Stanford University, \\
Stanford, CA 94305-4060, USA
\end{center}

\vskip 2cm

{\centerline{\large\bf Abstract}}
\begin{quotation}
\vskip -0.4cm
A theory of evolution of the universe requires both a
mutation mechanism and a selection mechanism. We believe
that both can be encountered in the stochastic approach to
quantum cosmology. In Brans--Dicke chaotic inflation, the
quantum fluctuations of Planck mass behave as mutations,
such that new inflationary domains may contain values
of Planck mass that differ slightly from their parent's.
The selection mechanism establishes that the value of
Planck mass should be such as to increase the proper
volume of the inflationary domain, which will then generate
more offsprings. This mechanism predicts that the effective
Planck scale at the end of inflation should be much larger
than any given scale in the model.
\end{quotation}
\newpage

The inflationary paradigm has two main features that makes it
very attractive. First, its classical evolution as an
exponential expansion of the universe, that solves
simultaneously many of the long standing problems of the
standard cosmological model, like the flatness, horizon and
homogeneity problems. On the other hand, its quantum evolution,
which provides a simple means of generating the required density
perturbations in the homogeneous background necessary for galaxy
formation. However, the same mechanism that
generates those ripples in the metric is responsible, at scales
much beyond our observable universe, for a very inhomogeneous
large scale structure of the universe \cite{LinBook}.

It is generally assumed that the gravitational dynamics of the
whole universe is correctly described by general relativity.
However, this seems to be a strong
assumption, extrapolating the description of the gravitational
phenomena at our local and low energy scales to the very large
scales beyond our observable universe. It is believed
that the theory of general relativity is just a low energy
effective theory of the gravitational interaction at the quantum
level. On the other hand, the low energy effective theory from
strings has the form of a scalar-tensor theory, with non-trivial
couplings of the string dilaton to matter.
The simplest scalar-tensor theory is Brans--Dicke (BD) theory of
gravity with a constant $\omega$ parameter. In this case, the
dilaton plays the role of the Brans--Dicke scalar field,
which acts like a dynamical gravitational
coupling, $M_{\rm p}^2(\phi) = {2\pi\over\omega} \phi^2$.
We will describe here the stochastic inflation associated with
such a theory or gravity \cite{JGB}.

One of the most fascinating features of inflation is the
self-reproduction of inflationary domains \cite{LinBook}. Due
to the existence of a horizon in de Sitter space, the
coarse-grained inflaton and dilaton fields undergo quantum
jumps of average amplitude $\delta\sigma = \delta\phi = H/2\pi$.
Such fluctuations act on the background fields as stochastic
forces \cite{GBLL}.
In their Brownian motion, the scalar fields may grow against their
drift force, very much like a Brownian particle in suspension
moving against the force of gravity, and take different values in
different causally disconnected inflationary domains.
Beyond a certain range of
values of the inflaton and dilaton fields, we enter the regime of
self-reproduction of the universe. Those few domains that jump
opposite to the classical trajectory contribute with a larger
proper volume and therefore dominate the physical space of the
universe. During the self-reproduction phase, the universe becomes
extremely inhomogeneous. As a consequence of the diffusion process,
there will always be domains that are still inflating and producing new
inflationary domains, while our own observable universe is just
the final evolutionary product of one of such domains \cite{LLM}.

A concept worth developing is the idea of Darwinian evolution
of fundamental constants from one inflationary domain to
another, as proposed by Linde \cite{PhysToday}. A theory of
evolution for the universe requires both a selection and a
mutation mechanism. We believe that both can be encountered in
the stochastic approach to quantum cosmology. In Brans--Dicke
chaotic inflation, the quantum jumps of Planck mass could
serve as a mechanism for mutation, such that new inflationary
domains may contain values of Planck mass that differ slightly
from their parent's. The selection mechanism establishes that
the value of Planck mass should be such as to increase the proper
volume of the inflationary domain, which will then generate more
``offsprings". It should be understood that this selection mechanism
works only if those values of the fundamental constants are
compatible with inflation. It is enough that new inflationary domains,
causally disconnected from each other thanks to the de Sitter
horizon, have different values of those constants. In the
case of Planck mass, the so called dilaton plays the role of the
fluctuating field. We will assume that certain ``phenotypical"
traits (a set of effective values for the fundamental constants)
will be selectively favored if, statistically, those inflationary
domains with such a set of values occupies the largest proper volume,
compatible with inflation \cite{evolution}.

In contrast to the case in general relativity \cite{LLM}, the
probability distribution that describes the diffusion process
in Brans-Dicke stochastic inflation never
becomes stationary for generic chaotic potentials \cite{GBLL}.
The distribution moves for ever towards larger values of the fields,
without crossing the Planck boundary. They constitute what we
called ``runaway" solutions.
There is a model independent prediction from the runaway character
of the probability distribution. In the case of an
inflaton of mass $m$, it dynamically predicts a very large Planck
mass compared to the only scale in the model, $m$. Furthermore,
since the amplitude of density perturbations at horizon scales goes
like $\delta\rho/\rho \sim m/ M_{\rm p}$, we have a prediction:
the larger is the Planck mass in a given inflationary domain, the
smaller is the amplitude of density perturbations. In the spirit
of Linde \cite{PhysToday}, the universe
evolves towards largest Planck mass and smallest amplitude of
density perturbations compatible with inflation, a prediction
that seems to agree well with observations. Our universe,
with our set of values for the fundamental constants, is the
offspring of one of such inflationary domains that started close
to Planck scale and later evolved towards the radiation and matter
dominated eras.

\end{document}